\documentclass[journal]{IEEEtran}
\usepackage[T1]{fontenc}
\usepackage{lmodern}
\usepackage{stackengine}

\usepackage{amsmath}

\usepackage{stackengine}

\usepackage{wrapfig}
\usepackage{caption}
\usepackage[framemethod=tikz]{mdframed}
\usetikzlibrary{calc}
\usepackage{kantlipsum}

\title{Modelling Underwater Acoustic Propagation using One-way Wave Equations}
\author{Indrakshi Dey, National University of Ireland, Maynooth, indrakshi.dey@mu.ie
}
\date{December 2021}

\begin{document}

\maketitle

\begin{abstract}
The primary contribution of this paper is to characterize the propagation of acoustic signal carrying information through any medium and the interaction of the travelling acoustic signal with the surrounding medium . We will use the concept of damped harmonic oscillator to model the medium and Milne's oscillator technique to map the interaction of the acoustic signal with the medium. The acoustic signal itself will be modelled using the one-way wave equation formulated in terms of acoustic pressure and velocity of acoustic waves through the medium. Using the above-mentioned concepts, we calculated the effective signal strength, phase shift and time period of the communicated signal. Numerical results are generated to present the evolution of signal strength and received signal envelope in underwater environment.
\end{abstract}

\section{Introduction}

A wide variety of applications like oil field exploration, underwater and underground environment monitoring, bio-geochemical process exploration etc. have made underground and underwater wireless communication systems a popular research area among both the academic and industry communities \cite{1}. However, underwater and underground environments pose challenges way different from air or space \cite{2}. Electromagnetic (EM) waves cannot be used to carry information in the above-mentioned application scenarios. This is because seawater is conductive and solid ground is absorptive and EM waves get attenuated exponentially over distance. Moreover, depending on the EM frequency range, EM waves do not travel through the actual channel. EM waves will only propagate through the upper layer of water or soil. The distance over which the EM waves can travel reliably decreases with the increase in the frequency of operation \cite{3}. 

In underwater and underground scenarios, engineers and physicists recommend acoustic waves for carrying information. Acoustic waves are mechanical waves whose direction of propagation is along the axis in which the acoustic field travels. They travel through adiabatic compression and rarefaction of the medium particles and move longitudinally along the medium. Their movement is guided by parameters like acoustic pressure, particle velocity, particle displacement and acoustic intensity \cite{4}. After being emitted from a source, acoustic waves travel in different directions sum of which gives rise to the acoustic pressure field. Depending on the environment through which the acoustic pressure is travelling, the family of acoustic waves get reflected, refracted and experience change in effective velocity, strength, phase shift and time period on its way to the destination \cite{5}. However, it is extremely challenging to mathematically describe the interaction of the propagating acoustic wave with the underwater and underground environments \cite{6}.

Different geographical areas and water-bodies are affected by different phenomena and parameters; and, therefore it is very difficult to propound a generalized modelling technique for acoustic wave propagation. Engineers have, so far, resorted to measurement campaigns, testing and validation effective only in a particular environment \cite{7}. Such characterization fails to model other environments thereby imposing a huge cost to bring in a little realism. 

\section{Background}

Acoustic wave propagation in underground scenario is affected by the reflection and attention coefficients of the soil, frequency of operation and velocity of acoustic waves through varying kinds of soil \cite{8}. On the other hand, in underwater scenario, acoustic signal is highly impacted by the temperature, pressure salinity, bubbles present in the water, ambient noise in the environment, change in velocity through interaction with the vibrating water molecules \cite{9}. Therefore, separate modelling approaches are used to characterize acoustic wave propagation through sea-surface, through areas close to the sea-floor and through middle of deep water-bodies. 

Sea-surface propagation models consider the impact of attenuation due to rough sea-surface and sea winds and the power spectral density can be computed as \cite{10},
\begin{equation}
 S(k) = \frac{\alpha}{2k^3}e^{-\frac{\beta g^2}{k^2 u^4}} \label{eq1}
\end{equation}
where $\alpha = 0.0081$, $\beta = 0.74$, $g = 9.82~m/s^2$ is the acceleration due to gravity, $u$ is the wind-speed at $19.5~m$ above the sea-surface and $k$ is the angular spatial frequency. Bathymetry is the most common model for acoustic wave propagation through regions close to the bottom of the sea. It is a model for multipath  scattering pattern caused by different sea-floor topologies. A common method for generating sea-floor topology is to compute the function \cite{11},  
\begin{equation}
\zeta(x) = R(x)~\frac{\zeta_{\text{max}}}{2} \bigg(\sin \bigg(- \frac{\pi}{2} + \frac{2\pi x}{L_h}\bigg) + 1\bigg)\label{eq2}
\end{equation}
where $\zeta(x)$ is the elevation at position $x$, $\zeta_{\text{max}}$ is the maximum underwater hill elevation, $L_h$ is the distance between adjacent peaks, and $R(x) \in (0,1]$ is the scaling function that generates uniform random number in different ranges of position. 

Propagation of acoustic waves through the central part of water-bodies is generally characterized using acoustic beams. BELLHOP \cite{12} is the most common software used for ray-tracing based on geometric beams. BELLHOP is a beam tracing model for predicting acoustic pressure fields in ocean environments. Several types of beams are implemented including Gaussian and hat-shaped beams, with both geometric and physics-based spreading laws. Energy spread over every acoustic beam is modeled using Gaussian intensity profile \cite{13} superposition of which can be used to estimate the total acoustic intensity at the receiver. However, there is no underlying mathematical framework that can describe the flow of acoustic signal through any environment.

This paper will introduce an entirely new approach using the concept of damped classical harmonic oscillators from physics to model the medium, interactions within the medium and their impact on the travelling signal. One-dimensional wave equation is used to model the travelling acoustic signal. As the acoustic signal propagates through the medium, its parameters like amplitude, phase, frequency and wavelength are modulated by its interaction with the particles within the medium. This paper will therefore use concept of parametric (Milne) oscillator for mapping the interaction between the medium and the propagating acoustic signal. Combining all the above concepts, it will be possible to quantify the propagating acoustic field strength, energy content within the field and energy dissipated due to interaction between the signal and the medium.

\section{Analytical Modelling}

\subsubsection{Modelling the Medium}

A linearly damped undriven harmonic oscillator is represented using \cite{14},
\begin{equation}
\ddot{x} + \beta \dot{x} + \omega^2 x = 0\label{eq3}
\end{equation}
where $x$ is the position, $\omega$ is the frequency of operation and $\beta$ is the coefficient of linear damping. If the oscillations are modified by varying some parameters of the system, the harmonic oscillator is referred to as parametric oscillator mathematically described by,
\begin{equation}
\ddot{x} + \beta(t) \dot{x} + \omega^2(t) x = 0\label{eq4}
\end{equation}
where the parameters $\omega$ and $\beta$ vary with time independent of the state of the oscillator, $\omega(t) > 0$ and $\omega(t) \to 1$ as $t \to \pm \infty$. Following the conventions in \cite{15} and \cite{16}, the limit for angular frequency $\omega(t)$ is set to 1 using constant time scaling. For the present scenario under consideration, where an acoustic wave carrying information travels through a medium (underwater acoustic propagation), we model the propagation environment using harmonic oscillators and the propagating acoustic signal using the acoustic wave equation. As the acoustic wave interacts with the surrounding water, the harmonic oscillators used to model the water movement can be modified to parametric oscillators. 

Parametric oscillators can be used to reflect the changes in the parameters of the surrounding water due to the interaction with the travelling acoustic waves. When no acoustic signal is travelling, the parametric oscillators representing the underwater environment can be visualized as unit frequency harmonic oscillator $(\omega = 1)$. The parametric oscillator again settles back to undriven state with unit frequency after the acoustic signal propagation has ended. However, over the time duration the signal exists, the fundamental solution to the harmonic oscillator gets modified and the solutions to the parametric oscillators can be visualized as a linear combination of the fundamental solutions. When the acoustic signal is interacting with the surrounding water, the change in the behavior of the propagation environment due to the acoustic signal is represented using a transition matrix.

\subsubsection{Modelling the Acoustic Signal}

We start with the one-dimensional acoustic wave equation given by \cite{17},
\begin{equation}
c^2 \frac{\partial^2 \mathbf{p}}{\partial x^2} = \frac{\partial^2 \mathbf{p}}{\partial t^2} \label{eq5}
\end{equation}
where $\mathbf{p}$ is the acoustic pressure, $x$ is the direction in which the acoustic signal is propagating and $c$ is the speed of sound in the medium. For example, $c = 343~m/s$ in air and $c = 1480~m/s$ in water. The equation in (\ref{eq5}) is generally solved in the form,
\begin{equation}
\mathbf{p} = f_1(x + ct) + f_2(x - ct) \label{eq6}
\end{equation}
where $f_1$ and $f_2$ are twice-differentiable functions each representing free travelling waves propagating in opposite directions (upwards and downwards along one dimension, for example $x$-direction). If we consider a sinusoidal acoustic wave carrying information, then the solution in (\ref{eq6}) modifies to,
\begin{equation}
\mathbf{p} = \alpha \cos k (x - ct) \label{eq7}
\end{equation}
where $\alpha$ is the amplitude of the acoustic signal wave, $k$ is the wave-number such that $k = \tilde{\omega}/c = 2\pi/\lambda$, $\tilde{\omega}$ is the frequency of the travelling acoustic signal and $\lambda$ is the wavelength of the signal. Rearranging (\ref{eq7}) in terms of $x$, we can obtain,
\begin{equation}
x = \frac{1}{k} \cos^{-1}\bigg(\frac{\mathbf{p}}{\alpha}\bigg) + ct\label{eq8}
\end{equation}
Inserting (\ref{eq8}) in (\ref{eq4}) we can obtain,
\begin{align}
&\ddot{\mathbf{p}}\frac{\mathbf{p}^2}{\alpha^2} - \mathbf{p}(\dot{\mathbf{p}})^2 + \beta(t) \dot{\mathbf{p}}\frac{\mathbf{p}^2}{\alpha^2} - \beta(t)ck\mathbf{p} \nonumber\\
&\quad - \omega^2(t)\mathbf{p}\cos^{-1}\bigg(\frac{\mathbf{p}}{\alpha}\bigg) - \omega^2(t)ctk\mathbf{p} = 0
\label{eq9}
\end{align}
Provided, $x = ct$ satisfies (\ref{eq9}), we can write the so-called Milne's equation \cite{18} where, $\mathbf{p}/\alpha = 1$; $\mathbf{p} = \alpha$ to obtain,
\begin{align}
\ddot{\mathbf{p}} - \mathbf{p}(\dot{\mathbf{p}})^2 + \beta(t) \dot{\mathbf{p}} - \beta(t)ck\mathbf{p} - \omega^2(t)ctk\mathbf{p} = 0
\label{eq10}
\end{align}
The Milne's equation in (\ref{eq10}) represents the time-evolution of the acoustic pressure field as it travels over a medium (air, water or underground) experiencing dynamic (time-dependent) variation in amplitude and phase. Equation (\ref{eq10}) can be solved numerically for $\mathbf{p}$ (the acoustic pressure field strength). From the Milne equation, we can also calculate the Milne energy which is equivalent to the effective receive signal strength. Correspondingly, effective phase shift and time period of the communicated signal cam also be calculated using the transition matrix formulated to reflect the interaction between the propagating acoustic signal and the medium.

\subsubsection{Modelling the Interaction between the Medium and the Acoustic Signal}

In order to solve for the acoustic pressure field $\mathbf{p}$ using the Milne's equation, we use the Lagrangian \cite{19, 20, 21} and the corresponding Hamiltonian formulation. This set of formulations can be used as they can be applied to the case of continuous media and fields with finite number of degrees of freedom. We start by formulating the Lagrangian of (\ref{eq10}) as,
\begin{align}
\mathcal{L} = \frac{1}{2} \dot{\mathbf{p}}^2 + \frac{\beta(t)ck}{2}\mathbf{p}^2 + \frac{\omega(t)^2ctk}{2}\mathbf{p}^2
\label{eq11}
\end{align}
In (\ref{eq11}), the Lagrangian density is formed through the difference between the kinetic energy density ($\frac{1}{2} \dot{\mathbf{p}}^2$) and the potential energy density ($\frac{\beta(t)ck}{2}\mathbf{p}^2 + \frac{\omega(t)^2ctk}{2}\mathbf{p}^2$). The Hamiltonian density can be obtained from (\ref{eq11}) using the concept of Euler-Lagrange equations as,
\begin{align}
\mathcal{H}(\dot{\mathbf{p}},p,t) &= \dot{\mathbf{p}}\frac{\partial \mathcal{L}}{\partial \dot{\mathbf{p}}} - \mathcal{L} \nonumber\\
&= \frac{1}{2} \dot{\mathbf{p}}^2 - \frac{\beta(t)ck}{2}\mathbf{p}^2 - \frac{\omega(t)^2ctk}{2}\mathbf{p}^2
\label{eq12}
\end{align}
The Hamiltonian is also the reflection of the difference between the potential and the kinetic energy content within the harmonic oscillator representing the medium. This energy content will remain constant before and after the duration over which the acoustic signal exists. Therefore we can now represent the Hamiltonian $\mathcal{H}$ as the asymptotic Milne energy $E_M$ or the effective signal strength at any point in space between the source and the destination of the acoustic signal.

We also introduce the Milne solution $\mathbf{q}$ which is asymptotic in anture and will satisfy the formulated Hamiltonian in (\ref{eq12}) such that the received signal strength (or the Milne energy) is bounded by $E_M \geq 1$. In turn, the acoustic pressure field $\mathbf{p}$ is non-stationary, asymptotic and varies with time. Since, $\mathbf{p}$ is time-varying, our considered parametric oscillators in (\ref{eq2}) also are time-dependent and their frequency of oscillation $\omega(t)$ and damping factor $\beta(t)$ are also time-dependent. Therefore, the corresponding Milne energy equation from the Hamiltonian can therefore be expressed as,
\begin{align}
E_M = \frac{1}{2} \dot{\mathbf{q}}^2 - \frac{\beta(t)ck}{2}\mathbf{q}^2 - \frac{\omega(t)^2ctk}{2}\mathbf{q}^2
\label{eq13}
\end{align}
where $\mathbf{q}$ is analogous to the acoustic pressure field $\mathbf{p}$.

Since acoustic pressure field are created through the summation of contributions of eigenrays at a particular point in space and time, $\mathbf{q}$ can be expressed as a function of time propagating along the $x,y$-plane. The amplitude of $\mathbf{q}$ can be written in the form,
\begin{align}
\mathbf{q} = \sqrt{\mathbf{q}^2_{\pm}\cos^2(t - \tau) + \mathbf{q}^2_{\mp}\sin^2(t - \tau)}
\label{eq14}
\end{align}
where $\tau$ is the effective time period of the acoustic signal propagating over the medium. Using (\ref{eq14}), derivative of $\mathbf{q}$, $\dot{\mathbf{q}} = 0$. Putting $\dot{\mathbf{q}} = 0$ back in (\ref{eq13}), we can express (\ref{eq13}) as,
\begin{align}
E_M = - \frac{\beta(t)ck}{2}\mathbf{q}^2 - \frac{\omega(t)^2ctk}{2}\mathbf{q}^2
\label{eq15}
\end{align}
which can be solved to obtain,
\begin{align}
\mathbf{q} &= \pm\imath\sqrt{2E_M/(\beta(t)ck + \omega(t)^2ctk)} \nonumber\\
\mathbf{q}(t) &= \pm\imath\sqrt{2E_M \cos(2t - \tau)/(\beta(t)ck + \omega(t)^2ctk)}
\label{eq16}
\end{align}
where $\imath = \sqrt{-1}$. 

The next step is to calculate the transition of the medium due to the interaction between the propagating acoustic field $\mathbf{p}$ and the medium. the transition of the oscillators representing the medium can be reflected through the formulation of the transition matrix $\mathbf{M}$. This transition matrix can be expressed as,
\begin{align}
\mathbf{M} = \mathbf{D}  \begin{pmatrix}
  \cos 2\delta & \mathbf{q}^2_{-}\sin 2\delta \\
  -\mathbf{q}^2_{+}\sin 2\delta & \cos 2\delta 
 \end{pmatrix} \mathbf{D};\nonumber\\
 \mathbf{D} = \begin{pmatrix}
  \cos\tau & -\sin\tau \\
  \sin\tau & \cos\tau
 \end{pmatrix}
\label{eq17}
\end{align}
where $\delta$ is the effective phase shift experienced by the acoustic signal. The equation for $\mathbf{M}$ can be simplified to form,
\begin{align}
\mathbf{M} = \begin{pmatrix}
  \cos\tau \cos\delta + \mathbf{q}^2_{-}\sin\tau \sin\delta & \mathbf{q}^2_{-}\sin\delta \cos\tau - \sin\tau \cos\delta\\
  \sin\tau \cos\delta - \mathbf{q}^2_{+}\sin\delta \cos\tau & \cos\delta \cos\tau + \mathbf{q}^2_{+}\sin\tau \sin\delta
 \end{pmatrix}
\label{eq18}
\end{align}
where 
\begin{align}
\mathbf{q}^2_{-} = \frac{2E_M \cos(2t - \tau)}{\beta(t)ck + \omega(t)^2ctk} \\
\mathbf{q}^2_{+} = - \frac{2E_M \cos(2t - \tau)}{\beta(t)ck + \omega(t)^2ctk}
\label{eq19}
\end{align}
For various values of the dynamical parameters, $E_M$, $\delta$ and $\tau$, we can have different values of $\mathbf{M}$. $\mathbf{M}$ varies with $E_M$ and correspondingly varies with $\mathbf{p}$ which in turn depends on the distance between the transmitter and the receiver.

\end{document}